\documentstyle[11pt,aaspp4]{article}

\begin{document}

\title{
	A maximum-likelihood method to improve
        faint source flux and color estimates
}
\author{
	David W. Hogg\altaffilmark{1,2} \&
	Edwin L. Turner\altaffilmark{2,3}
}
\altaffiltext{1}{
Theoretical Astrophysics, California Institute of Technology
}
\altaffiltext{2}{
School of Natural Sciences, Institute for Advanced Study,
Olden Lane, Princeton NJ 08540;
{\tt hogg@ias.edu}
}
\altaffiltext{3}{
Princeton University Observatory, Princeton University,
Peyton Hall, Princeton NJ 08544;
{\tt elt@astro.princeton.edu}
}

\begin{abstract}
Flux estimates for faint sources or transients are systematically
biased high because there are far more truly faint sources than
bright.  Corrections which account for this effect are presented as a
function of signal-to-noise ratio and the (true) slope of the
faint-source number-flux relation.  The corrections depend on the
source being originally identified in the image in which it is being
photometered.  If a source has been identified in other data, the
corrections are different; a prescription for calculating the
corrections is presented.  Implications of these corrections for
analyses of surveys are discussed; the most important is that sources
identified at signal-to-noise ratios of four or less are practically
useless.
\end{abstract}

% -----------------------------------------------------------------------------
\section{Introduction}

{\em Given a noisy photometric measurement of a very faint source,
what is the best estimate of its true flux?}

The best answer to this question is ``I don't know---integrate longer
to reduce the noise!''  However, in some cases, this is not possible.
For example, the ultra-deep images of the Hubble Deep Field (Williams
et al 1996) represent so much HST observing time that in practice they
cannot be much improved.  From the ground they cannot even in
principle be improved because any ground-based images significantly
deeper than existing ones (e.g., Djorgovski et al 1995; Metcalfe et al
1995; Smail et al 1995; Hogg et al 1997a, 1997b) would be totally
confusion-limited (e.g., Condon 1974).  As another example,
observations of transients, such as gamma-ray bursts or supernovae,
cannot be improved because they cannot be repeated, even in principle.

Given that in some cases deeper imaging is not an option, the reason
the question does not have a trivial answer is that the number counts
of faint sources tend to rise with decreasing flux, so more sources
are available for ``upscattering'' to a given measurement than are
available for ``downscattering.''  A familiar analogy is with
trigonometric parallaxes, where low signal-to-noise ratio measurements
are biased large, since given any observed parallax $\pi_o$ and
associated error, there is a finite range of true parallaxes $\pi$
consistent with it, but there are far more sources in the sky with
small parallaxes $\pi<\pi_o$ than large $\pi>\pi_o$.  The
``Lutz--Kelker'' corrections which account for this are easy to
compute and apply (Lutz \& Kelker 1973; Hansen 1979); they have been
essential in providing unbiased distances in astronomy.

There is a conceptually similar set of corrections for low
signal-to-noise ratio measurements of faint source fluxes.  In this
article, these corrections are computed and discussed.  As in the case
of parallaxes, the corrections depend on how the sources are selected,
and on the intrinsic distribution of the measured quantity, in this
case the true number-flux relation.  Unfortunately the number-flux
relation is not exactly known in most cases of interest, since the
faint source photometry is usually being performed in order to
determine this very relation!  Furthermore, in many cases of interest,
the correction presented here does not represent the largest source of
systematic error.  However, unlike the other sources of error, this
correction applies to all flux-selected sources, independent of
instrumentation or analysis technique.

One note of terminology: The systematic bias discussed here is often
improperly referred to as the ``Malmquist bias.''  The Malmquist bias
is the effect that in a flux-limited sample, there is a
larger-than-representative fraction of high-luminosity sources because
they can be seen to greater distances and hence over a larger volume
(Malmquist 1924; Mihalas \& Binney 1981).  It is due to the {\em
intrinsic\/} scatter in source luminosities. Malmquist bias is
removed, e.g., when one computes a luminosity function from star
counts.  It does not involve any kind of measurement error; it does
not go away if one obtains more precise photometry!  The bias
corrected-for here results from the {\em observational} scatter in
fluxes; the measurement errors.  It does indeed go away when the
fluxes are re-measured at much higher precision; it only needs to be
considered when low signal-to-noise data are being used.  What is
discussed in this paper is closely related to Eddington bias, the
effect of low signal-to-noise flux measurements on faint source
number-magnitude relations.  Statistical corrections to observed
number-magnitude relations are computed by Eddington (1913); flux
corrections for individual survey sources are computed here.

% -----------------------------------------------------------------------------
\section{Computation of corrections}
\label{sec:corr0}

Consider the simplest case, in which a source is being photometered in
the image in which it was first detected.  That is, it is being
measured in the data in which it was selected.  The likelihood
$p(S|S_o)$ (probability per unit flux) that a source has true flux $S$
given that it is observed to have flux $S_o$ is related to the
likelihood $p(S_o|S)$ that it is observed to have $S_o$ when it has
$S$ by Bayes's theorem
\begin{equation}
p(S|S_o)\propto p(S_o|S)\,p(S)
\end{equation}
where a proportionality is used because the normalization is being
ignored (for now) and $p(S)$ (probability per unit flux) is the true
underlying distribution of fluxes, given by the (true, not observed)
number-flux relation.  If the number of sources $N(<m)$ brighter than
magnitude $m$ as a function of $m$ is a power law
\begin{equation}
2.5\,\frac{d\log N}{dm}=-\frac{d\log N}{d\log S}=q
\end{equation}
then the conditional probability becomes
\begin{equation}
p(S|S_o)\propto\left(\frac{S}{S_o}\right)^{-(q+1)}
 \,\exp\left[-\frac{(S-S_o)^2}{2\sigma^2}\right]
\label{eq:probgauss}
\end{equation}
where it is assumed that the observational error is
gaussian-distributed and $\sigma$ is the uncertainty in the observed
flux $S_o$, or $S_o/\sigma$ is the signal-to-noise ratio $r$.
Figure~\ref{fig:plotbias} shows these likelihood curves for
number-flux exponent $q=2.0$, $1.5$, $1.0$ and $0.5$ and
signal-to-noise ratios $r=3$, $5$ and $10$.  This Figure demonstrates
that measurements at a signal-to-noise ratio of three do not strongly
constrain the true flux, whatever the slope of the number counts, but
particularly if the counts have the Euclidean\footnote{Another note on
terminology: What is called the ``Euclidean'' slope really ought to be
called the ``no-evolution, non-expanding'' slope, because even in a
Euclidean space, the number counts have $q\neq 1.5$ at large distance
if either the Universe is expanding or the sources are evolving.}
slope of $q=1.5$ (or greater).  It is worth emphasizing that the above
equation and the curves plotted in Figure~\ref{fig:plotbias} are
essentially identical to those computed for parallax corrections (Lutz
\& Kelker 1973; Hansen 1979) except that the parallax corrections are
computed for only one particular exponent value.

If the flux measurement was unbiased, the peak in the likelihood
function $p(S|S_o)$ would be at $S/S_o=1$.  However, taking the
derivative $dp/dS$, it is found that the maximum-likelihood true flux
$S_{\rm ML}$ is in fact
\begin{equation}
\label{eq:correction}
\frac{S_{\rm ML}}{S_o}=\frac{1}{2}+\frac{1}{2}\,\sqrt{1-\frac{4\,q+4}{r^2}}
\end{equation}
where $q$ is the number-magnitude exponent defined above and $r$ is
the signal-to-noise ratio.  There is no finite maximum-likelihood
value at all if $r^2<4\,q+4$; an example is the $q=1.5$, $r=3$ curve in
Figure~\ref{fig:plotbias}.  {\em The above equation specifies a
correction which should in principle be applied to all flux
measurements in a flux-limited sample.}  When the signal-to-noise is
good enough ($r^2\gg 4\,q+4$) the correction can be approximated as
\begin{equation}
\frac{S_{\rm ML}}{S_o}\approx 1-\frac{q+1}{r^2}
 \;\;\; {\rm when} \;\;\; r^2\gg 4\,q+4
\end{equation}
or in terms of the magnitude correction $\Delta m\equiv m_{\rm ML}-m_o$
\begin{equation}
\Delta m\approx\frac{1.086\,q+1.086}{r^2}
 \;\;\; {\rm when} \;\;\; r^2\gg 4\,q+4
\end{equation}

Things change slightly if the likelihood is computed in the magnitude
(i.e., log flux rather than flux) domain; after all,
maximum-likelihood techniques are sensitive to the ``metric'' of the
space in which the likelihood is computed.  When computed purely in the
magnitude or log-flux domain, the correction is
\begin{equation}
\Delta m\approx\frac{1.086\,q+2.171}{r^2}
 \;\;\; {\rm when} \;\;\; r^2\gg 4\,q+8
\end{equation}
The fact that the solution depends on the type of data space (log or
linear) demonstrates that the specific value of the correction is not
completely specified, because it only provides a ``best guess'' (a
subjective estimate) for the true flux.  It is worthy of note that the
Lutz-Kelker parallax corrections are similarly subjective, as are
essentially all statistical estimators.

More robust than maximum-likelihood estimates are confidence
intervals, because these do not depend on the choice of ``metric.''
Confidence intervals are found by integrating the likelihood curves.
Unfortunately, the areas under the curves shown in
Figure~\ref{fig:plotbias} do not converge; the likelihood
distributions are not normalizable!  This non-normalizability comes
from the divergence of $p(S|S_o)$ as $S\rightarrow 0$ (not visibile in
some of the curves in Figure~\ref{fig:plotbias} simply because at high
$r$ the divergence happens at very small $S/S_o$).  There are two
respects in which this divergence or non-normalizability is
unphysical: First, there cannot be an infinite number of sources in
the visible Universe; there aren't even an infinite number of {\em
atoms\/} in the Universe!  Secondly, most ultra-deep images of the
sky, including the HDF, are close to their confusion limits, beyond
which the observed number counts have to ``cut off'' no matter how
much integration time is employed.  Neither of these effects can be
simply taken into account in general; they depend on the data quality
and the sources under study.

The equations in this Section have assumed that observational errors
are gaussian-distributed, which is not true for all photometric
measurements.  The equations are easily generalized (although they do
not necessarily remain analytic) with the gaussian in
equation~(\ref{eq:probgauss}) replaced by whatever error distribution
is appropriate for the measurement in question.

% -----------------------------------------------------------------------------
\section{An empirical test}

The correction can be tested with any imaging data in which the
number-flux relation is known.  Here, the HST HDF data in the F606W
($0.6~{\rm \mu m}$) bandpass are used.  Noise was added to the
$1024\times 1024$ ``Version 2'' mosiacs of the HST images of the HDF
(Williams et al 1996) to make the pixel-to-pixel sky noise ten times
as bad as in the original mosaics.  The higher-noise mosaics will be
referred to as the ``bad'' images and the originals as the ``good''
images.  A catalog of sources was chosen in the bad images down to
very faint levels using the ``SExtractor'' source detection package
(Bertin \& Arnouts 1996) in essentially its default mode: Smooth with
a 2-pixel FWHM triangular filter and select take sources whose central
pixel in the smoothed image is above a given threshold.  These sources
were then photometered with the NOAO ``IRAF'' software in matched
0.16~arcsec (2~pixel) diameter apertures in both the bad and good
images.  The bad/good flux ratios are plotted against signal-to-noise
ratio in Figure~\ref{fig:scatter} along with the expected correction
computed with equation~(\ref{eq:correction}) and the (known) count
slope $q=0.5$ (Williams et al 1996).  The correction does very well
down to signal-to-noise ratios $r\approx 3$.  At $r<4$, a significant
number of spurious (zero-flux in the good image) sources start to
appear.  Figure~\ref{fig:scatter} shows, as with the Lutz-Kelker
corrections, that the corrections are on the same order as the
intrinsic scatter due to measurement error, so some sources have
under- rather than overestimated fluxes.  However the correction is
still necessary if an unbiased estimator of the true flux is desired.

% -----------------------------------------------------------------------------
\section{Changing the selection technique}

The next case to consider is photometry of a source in one image (say
$I$-band) after it is detected (and its position is known) in another
image (say $V$-band).  In this case, Bayes's theorem is still used but
for $p(S)$, the true distribution of $V-I$ colors is used rather than
the $I$-band number counts.  Actually, rather than the color
distribution, it is better to think of the conditional $I$-band flux
distribution $p(S^{(I)}|S^{(V)})$ (probability per unit flux) given
that the $V$-band flux $S^{(V)}$ is known (in what follows it is assumed
that the $V$-band detection is at very high signal-to-noise ratio so
the $V$-band flux is well known).  Because, unlike the number counts,
these conditional distributions are not generally power laws, the flux
correction depends not only on the shape of the distribution but where
in the distribution the observed flux $S_o$ lies.

Fortunately, when the signal-to-noise ratio $r$ is large enough, it is
possible to linearize Bayes's formula around the observed flux
$S^{(I)}_o$ so only the local power-law slope
\begin{equation}
Q\equiv -\left.\frac{d\log p(S^{(I)}|S^{(V)})}{d\log S^{(I)}}\right|_{S^{(I)}_o}
\end{equation}
of the distribution of source fluxes is important.  The likelihood
function (probability per unit flux) for the true $I$-band flux
$S^{(I)}$ given the observed flux $S^{(I)}_o$ and the known $V$-band
flux $S^{(V)}$ is then
\begin{equation}
p(S^{(I)}|S^{(I)}_o,S^{(V)})\propto\left(\frac{S^{(I)}}{S^{(I)}_o}\right)^{-(Q+1)}
 \,\exp\left[-\frac{(S^{(I)}-S^{(I)}_o)^2}{2\sigma^2}\right]
\label{eq:igivenv}
\end{equation}
which leads to the maximum-likelihood correction
\begin{equation}
\frac{S^{(I)}_{\rm ML}}{S^{(I)}_o}\approx 1-\frac{Q+1}{r^2}
 \;\;\; {\rm when} \;\;\; r^2\gg 4\,Q+4
\end{equation}
or in terms of the magnitude correction $\Delta m\equiv m_{\rm ML}-m_o$
\begin{equation}
\Delta m\approx\frac{1.086\,Q+1.086}{r^2}
 \;\;\; {\rm when} \;\;\; r^2\gg 4\,Q+4
\end{equation}
Note that the correction can be positive or negative, depending on the
sign of the local slope $Q$.

Corrections applicable when more complicated selection procedures have
been used can be computed in analogous ways.

% -----------------------------------------------------------------------------
\section{Summary and discussion}

Maximum-likelihood corrections for faint source flux measurements have
been computed for the case in which the sources are measured at low
signal-to-noise in the data in which they were originally selected.
It is found that since the number-flux relation tends to be rising at
the faint end, the low signal-to-noise flux measurements are usually
overestimates of the true flux.  At signal-to-noise ratios $r<4$, flux
measurements (of this type---i.e. in the data in which the sources
were selected) are almost meaningless because they are consistent with
almost any true flux between zero and the measured value.

The bias considered here tends to ``steepen'' measured number-flux
relations at the faint end; i.e. the measured $d\log N/d\log S$ is
more negative than the true value because the very numerous faint
sources are scattered up to brighter levels.  This effect is only
significant at very faint levels, where it is usually mitigated or in
fact canceled out by incompleteness.  The best way to correct measured
number-flux relations for both the flux bias and incompleteness is to
perform full completeness simulations, which, if done correctly, will
account for both effects simultaneously (e.g., Smail et al 1995; Hogg
et al 1997b), and of coures a full accounting for all systematic
errors requires detailed modeling of every stage in the observing and
analysis procedures.  Although the corrections presented here do not
comprehensively account for most of these systematic biases, they are
very general, improving flux estimates for individual sources
independent of observational technique.

These corrections ought to be applied to the source fluxes at the
faint end of the catalogs from all huge (and therefore
difficult-to-improve-upon) surveys, such as the Palomar Observatory
Sky Surveys, the Infrared Astronomical Sattelite (IRAS) survey and
from future huge surveys such as the 2-Micron All Sky Survey and the
Sloan Digital Sky Survey.  In fact the IRAS catalogs were corrected at
the faint end for some related biases but not this bias per se
(Beichman et al 1988).  Also, all transients discovered at low
signal-to-noise ratios in transient searches, such as faint gamma-ray
bursts, of which no additional measurements can be made after the
fact, should have these corrections applied.

% -----------------------------------------------------------------------------
\acknowledgements We thank Roger Blandford, John Gizis, Gerry
Neugebauer, Neill Reid and Yun Wang for useful discussions, and the
HDF team for planning, taking, reducing and calibrating the HDF data.
Some financial support was provided by NSF grant AST--9529170.

% -----------------------------------------------------------------------------
\newpage
\plotone{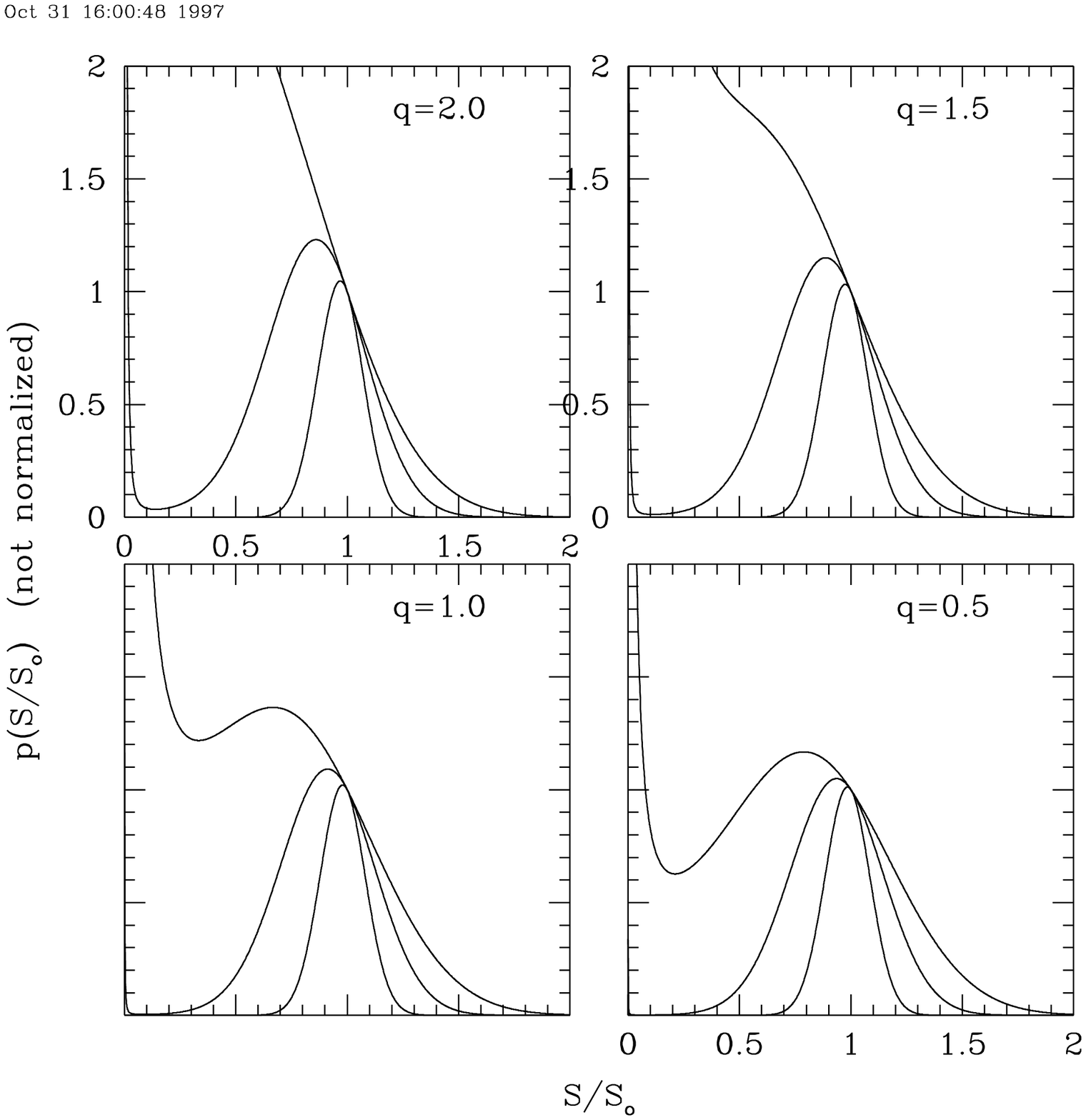}
\figcaption[plotbias.eps]{ Likelihood curves for several
number-magnitude exponents $q$ and signal-to-noise ratios $r=3$
(worst), 5 (middle) and 10 (best).
\label{fig:plotbias}
}

\newpage
\plotone{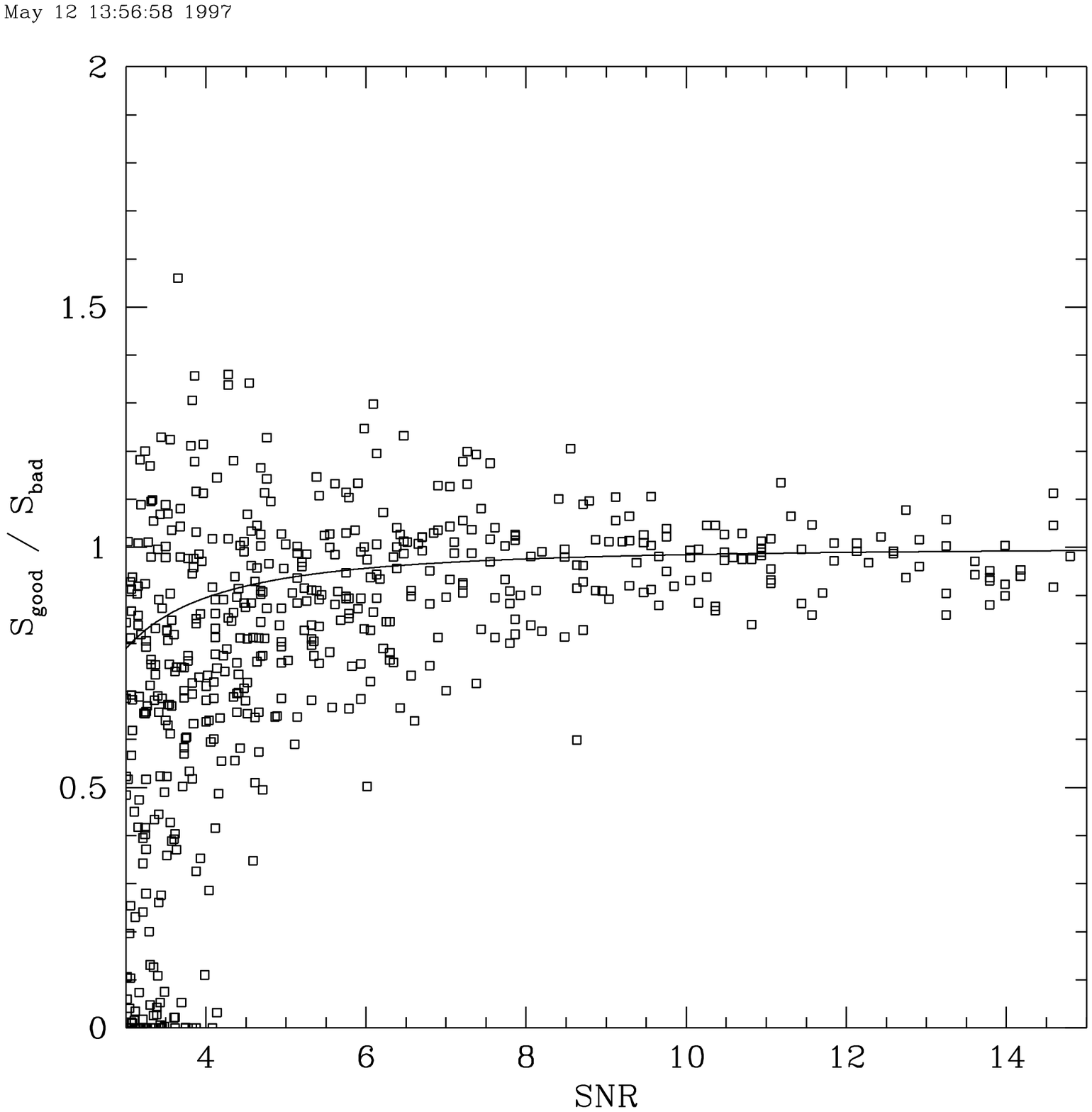}
\figcaption[scatter.eps]{ The ratios of flux measurements in the
``bad'' image, in which sources in the HDF were detected, to the flux
measurements in the ``good'' image.  The bad image is simply the good
image plus additional noise (see text).  The solid line is the
expected ratio $S_{\rm ML}/S_o$ of the maximum-likelihood flux to the
observed flux given by equation~(\ref{eq:correction}) for
number-magnitude exponent $q=0.5$ (see text).
\label{fig:scatter}
}

\end{document}